\documentclass[twocolumn,showpacs,preprintnumbers,superscriptaddress,amsmath,amssymb]{revtex4}
\usepackage{dcolumn}% Align table columns on decimal point
\usepackage{bm}% bold math
\usepackage{amsmath}
\usepackage{multirow}
\usepackage{graphicx}

 \newcommand{\mean}[1]{\mbox{$\langle{#1}\rangle$}}

\begin{document}

\title{Three-Dimensional Analysis of Wakefields Generated by \\
Flat Electron Beams in Planar Dielectric-Loaded Structures}

\author{D. Mihalcea}
\affiliation{Northern Illinois Center for
Accelerator \& Detector Development and Department of Physics,
Northern Illinois University, DeKalb IL 60115,
USA}
\author{P. Piot} \affiliation{Northern Illinois Center for
Accelerator \& Detector Development and Department of Physics,
Northern Illinois University, DeKalb IL 60115,
USA}\affiliation{Accelerator Physics Center, Fermi National
Accelerator Laboratory, Batavia, IL 60510, USA}
\author{P. Stoltz} \affiliation{Tech-X Corporation, Boulder, CO 80303, USA}

\date{\today}
%%\maketitle

\begin{abstract}
An electron bunch passing through dielectric-lined waveguide generates $\check{\mbox{C}}$erenkov radiation that can result in high-peak axial electric field suitable for acceleration of a subsequent bunch. Axial field beyond Gigavolt-per-meter are attainable in structures with sub-mm sizes depending on the achievement of suitable electron bunch parameters. A promising configuration consists of using planar dielectric structure driven by flat electron bunches. In this paper we present a three-dimensional analysis of wakefields produced by flat beams in planar dielectric structures thereby extending the work of Reference [A.~Tremaine,  J.~Rosenzweig, and P.~Schoessow, Phys. Rev. E {\bf {56}}, No. 6, 7204 (1997)]  on the topic. We especially provide closed-form expressions for the normal frequencies and field amplitudes of the excited modes and benchmark these analytical results with finite-difference time-domain particle-in-cell numerical simulations. Finally, we implement a semi-analytical algorithm into a popular particle tracking program thereby enabling start-to-end high-fidelity modeling of linear accelerators based on dielectric-lined planar waveguides. 
\end{abstract}

\keywords{wakefield, dielectric-loaded waveguide, field gradient, 
dispersion equation, Hertzian vector potential.}

\pacs{29.27.-a,41.20.Jb,41.60.Bq}
\maketitle

\section{INTRODUCTION}
Next generation multi-TeV high-energy-physics lepton accelerators are likely to be based on non-conventional acceleration techniques given the limitation of radio-frequency (rf) normal-conducting~\cite{Wuensch} and super-conducting~\cite{Aune} structures. Non-conventional approaches based on the laser plasma wakefield accelerator have recently demonstrated average energy gradients one order of magnitude higher than those possible with state-of-the art conventional structures~\cite{Leemans}. The applicability of laser-driven techniques to high-energy accelerators is currently limited as attaining luminosity values similar to those desired at the International Linear Collider would demand a laser with power approximately four orders of magnitude larger than the most powerful lasers currently available~\cite{Melissinos}. Another class of non-conventional accelerating techniques includes beam-driven methods which rely on using wakefields produced by high charge ÒdriveÓ bunches traversing a high-impedance structure to accelerate subsequent ÒwitnessÓ bunches~\cite{Voss}. Such an approach has the advantage of circumventing the use of an external power source and can therefore operate at mm and sub-mm wavelengths. Structures capable of supporting wakefield generation include plasmas~\cite{Blumenfeld}, and dielectric-loaded waveguides~\cite{Ng}. The possible use of plasma-wakefield accelerators (PWFAs) as the backbone of a a multi-TeV electron-positron linear collider is limited by plasma ionsÕ motion due to the intense electromagnetic field of the bunch~\cite{RosenPWFA}. Dielectric wakefield accelerators (DWFAs) are not prone to similar limitations. 

In this paper we concentrate on the colinear DWFA. In such a configuration a highly-charged drive bunch propagates through a dielectric-lined waveguide (DLW) and excites an electromagnetic wake~\cite{Chang,Rosing,Ng}. A delayed witness bunch moving on the same path as the drive bunch can experience an accelerating field. 

Recent experiments~\cite{Thompson} confirm that DLW can support accelerating fields in excess of a GV/m thereby making DWFA  a plausible candidate for the next generation high-energy-physics linear accelerators~\cite{GaiAAC10} or compact short-wavelength free-electron lasers~\cite{ZolentsPriv}.

In cylindrically-symmetric DLW structures, the electric field amplitude of the wakefield is approximately inverse proportional to the aperture radius. Given the linear charge-scaling of the field, peak electric field amplitude of the order of Gigavolt-per-Meter  can be obtained by either using high-charge drive beams ($>100$~nC) in mm-sized DLW structures or by focusing sub-nC bunches in micron-sized DLW structures~\cite{Tremaine,Rosing,Park}.

To date, cylindrically-symmetric DLW's have been extensively studied both theoretically \cite{Rosing,Bolotvskii,Liu,Sotnikov} and experimentally \cite{Power,Gai,Fang,Jing1}. Since the angular divergence of the beam also increases during focusing it is hard to maintain a low transverse size of the beam over a long propagation distance. Therefore, the design of the DLW structures must compromise between a small transverse size to maximize the intensity of the wakefield, and a longer interaction length to maximize the energy gain of the test charge.
An appealing solution consists of using flat electron beams ($\sigma_{x} \gg \sigma_{y}$) passing through slab-geometry DLWs~\cite{Tremaine,
Xiao,Jing2,Wang}. This possibility has become more attractive since the recent advances toward generating  flat beams directly in photoinjectors~\cite{Brinkmann,Piot}. 

In the following Sections we present an analysis of the generation of wakefields in rectangular DLW structures excited by drive beams with arbitrary three-dimensional charge distributions. The main goal of this paper is to extend the formalism introduced in Ref.~\cite{Tremaine} by including all types of modes excited in a planar DLW. This paper is pedagogical in the sense that it is about the method of derivation rather than the results, which have been derived in previous papers, most extensively in Ref.~\cite{Jing2}. However, this paper provides close-form formulae for the eigenfrequencies and electromagnetic fields excited in planar DLWs. These analytical results are benchmarked against three-dimensional finite-difference time-domain (FDTD) simulations. Finally, the model is implemented in a popular particle-in-cell (PIC) beam dynamics program. The latter provides a fast and high-fidelity model enabling start-to-end simulation of DLW-based linear accelerators. 

\section{WAKEFIELD GENERATION\label{secwake}}

\begin{figure}[htb]
\centering
\includegraphics*[width=80mm]{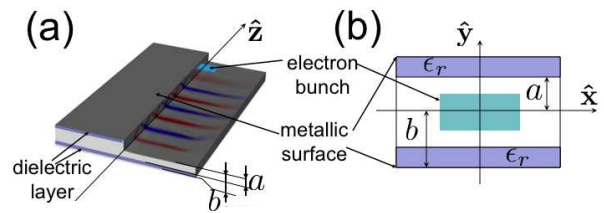}
\caption{Overview of the DLW structure geometry (left) and transverse cross section (right). A rectangular-shaped drive beam is displayed in blue.}
\label{DLW}
\end{figure}

The geometry of the problem analyzed in this paper is depicted in Fig.~\ref{DLW}.  Transient effects resulting from the injection of the electron bunch in the structure are not included (the structure is assumed to be infinitely long) and the drive bunch is taken to be ultra-relativistic with its Lorentz factor $\gamma$ much greater than unity. 
The building block of a real drive bunch is assumed to consist
of a linear charge distribution oriented along the x-axis which
moves in the $z$-direction with velocity $v$ and has offset $y_0$ in the
vertical direction. To simplify, it is also assumed that the linear
charge distribution is symmetric in the $x$-coordinate and it vanishes
at the ends of the dielectric structure.
%The longitudinal motion of the drive bunch is assumed to be at a constant
%speed $v$ and the origin of the $z$-axis is chosen such that $z=vt$ where $t$ is the time. %
%The bunch charge distribution is taken to be symmetric with respect to the vertical $y$-axis  
%[$\rho(x,y,z) = \rho(-x,y,z)$] and vanishes  at the dielectric
%and metallic boundary surfaces [$\rho(-L_{x}/2,y,z)=\rho(L_{x}/2,y,z)=\rho
%(x,a,z)=\rho(x,-a,z)=0$].  The bunch is modeled has a superimposition of line of charges with finite extension in horizontal direction and with an offset $y_0$ in the vertical direction.
Under these assumptions the charge distribution can be written as a Fourier series  in the horizontal direction~\cite{Tremaine}
\begin{equation}
\rho(x,y,z)=\sum_{m} \lambda_m  \cos(k_{x,m}x) 
\delta(y-y_0) \delta (z-vt) , 
\label{charge}
\end{equation}

\noindent where $\lambda_m$ is a constant and each term is indexed by the integer $m=0,1,\cdots$ defined such that $k_{x,m} \equiv  (2m+1) \frac{\pi}{L_x}$. In the charge-free vacuum region the electromagnetic wakefield satisfies the wave equation:
\begin{equation}
\left( \nabla^2 - \frac{1}{c^2} \frac{\partial^2}{\partial^2 t}
\right) \mathbf{E}=0. 
\label{wave}
\end{equation}
In this paper we consider only the propagating modes with harmonic longitudinal and temporal dependencies of the form $\mathbf{E}(x,y,z,t)=\mathbf{E}(x,y) e^{i(\omega t - k_z z)}$~\cite{Tremaine,Wang} where $k_z=\mathbf{k}.\mathbf{\hat{z}}$. 
In addition to the propagating-mode solutions, Eq.~\ref{wave} when subject to the boundary conditions also admits evanescent modes with imaginary eigenfrequencies $\omega$~\cite{Wang}. Since these evanescent modes are present over short distances behind the drive bunch, they do not contribute to the beam dynamics of a subsequent witness  bunch. These short-range fields are consequently ignored in the remaining of this paper.

In the ultra-relativistic regime the synchronism condition is achieved
because the wakefield phase velocity 
$v_\varphi \equiv \frac{\omega}{k_z} = v \approx c$ is very close to
the velocity of the witness beam. Since the  transverse wave vector $k_{\bot}\equiv\sqrt{k^2-k_z^2} \simeq k/\gamma = 0$  the wave equation~\ref{wave} can be written only in terms of the partial derivatives with respect to the transverse coordinates:
\begin{equation}
\nabla_{\bot}^{2} \mathbf{E} (x,y) = 0. 
\label{vac}
\end{equation}
The symmetry of the drive-beam charge distribution and the boundary conditions at $x =\pm L_{x}/2$, determine  (up to constants) the analytical expression of the fields. In the vacuum region the electric field components  are simple combinations of trigonometric and hyperbolic functions:
\begin{eqnarray}
E_x &\propto &\sin(k_{x,m} x) \cosh(k_y y), \nonumber \\
E_y &\propto &\cos(k_{x,m} x) \sinh(k_y y), \\
E_z &\propto &\cos(k_{x,m} x) \cosh(k_y y). \nonumber 
\label{struct1}
\end{eqnarray}
The axial field $E_z$ associated with this set of solution is symmetric with respect to the horizontal axis [i.e. $E_z(x,-y,z)=E_z(x,y,z$)] we henceforth refer this set to as  ``monopole" modes. Similarly, a set of fields with an antisymmetric axial field [$E_z(x,-y,z)=-E_z(x,y,z$)] is obtained  by substitution of $\sinh (k_{y} y)$ with $\cosh (k_{y} y)$ and vice versa. This latter set is termed as ``dipole" modes in the remaining of this paper. 

Inside the dielectric, the transverse wave number cannot be neglected

\begin{eqnarray}
k_{\bot}^2 \equiv k_{x,m}^{2} + k_{y}^{2} \approx \frac{\omega^2}{c^2}
\left(\epsilon_r - 1 \right) > 0, 
\label{kp}
\end{eqnarray}

\noindent where $\epsilon_r$ is the relative electric permittivity of the dielectric medium.  The boundary conditions at $y = \pm b$ and at  $x = \pm L_x / 2$ determine the trigonometric form of the field expressions.
Inside the dielectric region, there is no distinction between monopole and dipole modes and the electric field components are given by
\begin{eqnarray}
E_x &\propto &\sin(k_{x,m} x) \cos[k_y(b-y)] ,\nonumber \\
E_y &\propto &\cos(k_{x,m} x) \sin[k_y(b-y)] , \\
E_z &\propto &\cos(k_{x,m} x) \cos[k_y(b-y)] . \nonumber
\label{struct2}
\end{eqnarray}
The corresponding expressions for the magnetic field in the vacuum and dielectric regions can be easily obtained following a similar prescription.

Inspection of the $z$-component of the fields indicates that  the normal modes cannot be categorized in the usual transverse electric  or transverse magnetic sets. The reason is that the separation surface
between the dielectric and vacuum regions is in the $x-z$ plane unlike the case of the uniformly-filled waveguides where this surface is in the transverse $x-y$ plane. \\
Therefore, it is natural to categorize the modes depending on the orientation of the fields with respect
to the dielectric surface. Following the definition introduced in  Ref.~\cite{Collin} we classify the mode as Longitudinal Section Magnetic (LSM) and Longitudinal Section Electric (LSE) modes corresponding respectively to the case when the magnetic and electric field component perpendicular to the dielectric surface vanishes. In the configuration shown in Fig.~\ref{DLW},  LSE and LSM modes correspond respectively to $E_y = 0$ and $H_y = 0$ at the vacuum-dielectric interface. 
 
\section{DISPERSION EQUATIONS \& Electromagnetic wakefields}

\subsection{Longitudinal Section Magnetic (LSM) modes}
To obtain the normal mode frequencies it is convenient to use the 
hertzian potential vector method~\cite{Collin}. In the source-free
region, magnetic and electric fields can be expressed in terms of
the vector potential $\mathbf{\Pi}$:

\begin{eqnarray}
\label{vhp_lsm1}
\mathbf{H} &=& i \omega \epsilon \boldsymbol{\nabla} \times 
\mathbf{\Pi},\\
\label{vhp_lsm2}
\mathbf{E} &=& k^2 \mathbf{\Pi} + \boldsymbol{\nabla} 
( \boldsymbol{\nabla}  \cdot \mathbf{\Pi}) , 
\end{eqnarray}

\noindent where $\mathbf{\Pi}$ is the solution of the
homogeneous Helmholtz equation in the vacuum region.
To determine the frequencies and the fields for the modes when 
$H_y = 0$ (LSM) it is convenient to choose 
$\mathbf{\Pi} \equiv \mathbf{\Pi_e} =  \psi_e (x,y) e^{i(\omega t - k_z z)}\hat{\mathbf{y}}$ where the
the scalar function $\psi_e (x,y)$ must be a solution of  $(\nabla_{\bot}^{2} + k_{\bot}^{2})\psi_e = 0$. 
From Eqs.~\ref{vhp_lsm1} and~\ref{vhp_lsm2} all fields can be expressed in terms of the unknown function $\psi_e (x,y)$:

\begin{eqnarray}\label{lsmfields}
&E_x = \frac{\partial^2 \psi_e}{\partial x \partial y}  &\hspace{5mm} H_x = -\epsilon k_z^{2} c \psi_e \nonumber \\
&E_y = k^2 \psi_e + \frac{\partial^2 \psi_e}{\partial y^2}    &\hspace{5mm} H_y = 0  \\
&E_z = -i k_z \frac{\partial \psi_e}{\partial y}  &\hspace{5mm}  H_z = i \epsilon k_z c \frac{\partial \psi_e} {\partial x}  \nonumber
\end{eqnarray}

The expression for $\psi_e$ that satisfies Eqs.~\ref{lsmfields}, 
\ref{struct1}, and~\ref{struct2} is given by:

\begin{eqnarray}
\label{psie}
\psi_e =
\begin{cases}
A \cos(k_{x,m} x) \sinh(k_{x,m} y), & 0 < y < a \\
B \cos(k_{x,m} x) \cos \left[ k_y(b-y) \right], & a < y < b
\end{cases}
\end{eqnarray}

where $A$ and $B$ are constants. The boundary conditions
completely determine the eigenfrequencies of the LSM modes.
Since $H_x$ ($\propto \epsilon \psi_e$) and $E_z$ ($\propto \frac{\partial \psi_e}{\partial y}$) are
continuous at $y = a$ the constants $A$ and $B$ can be eliminated giving the dispersion equation

\begin{eqnarray}
\label{disp_lsm}
\coth(k_{x,m} a) \cot \left[ k_y(b-a) \right] = \frac{k_y}{\epsilon_r  k_{x,m}}. 
\end{eqnarray}

Therefore, for each discrete value of $k_{x,m}$ there is  an infinite set of discrete $k_{y,n}$ values where $n$ is an integer. So, the eigenfrequencies are indexed by the integer couple $(m,n)$ and verify 
\begin{equation}
k_{x.m}^2 + k_{y,n}^2 = \frac{\omega_{m,n}^2}{c^2} (\epsilon_r -1). 
\end{equation} 
It is worthwhile noting that the boundary conditions 
do not completely determine
either the function $\psi_e$ or the fields. The reason being that, up to this point, the field
source terms $(\rho, \mathbf{j})$ were not taken into account 
although their symmetries were invoked. Still, it is straightforward 
to show that all fields (and also $\psi_e$) associated to a certain mode
depend on a common normalization constant, the amplitude $E_{0;m,n}$, 
which remains to be determined.
The expressions of the fields are given by 

\begin{widetext}
\begin{align}
\label{Ez}
E_{x,m,n}=&
\begin{cases}
-\frac{i E_{0;m,n} k_{x,m}}{k_z} \sin(k_{x,m} x) \cosh(k_{x,m} y) & 0 < y < a \\
-\frac{i E_{0;m,n} k_{x,m}}{k_z} \frac{\cosh(k_{x,m} a)}{\sin \left[ k_{y,n}(b-a) \right]}
\sin(k_{x,m} x) \sin \left[ k_{y,n}(b-y) \right] & a < y < b
\end{cases}
\nonumber \\ 
\nonumber \\
E_{y,m,n}=&
\begin{cases}
\frac{i E_{0;m,n}}{k_{x,m} k_z} (k_{x,m}^2+k_z^2) \cos(k_{x,m} x) \sinh(k_{x,m} y) 
& 0 < y < a \\
\frac{i E_{0;m,n} \cosh(k_{x,m} a)}{k_{y,n} k_z \sin\left[k_{y,n}(b-a)\right]} 
(k_z^2+k_{x,m}^2)
\cos(k_{x,m} x) \cos \left[ k_{y,n}(b-y) \right] & a < y < b
\end{cases}
\nonumber \\ 
\nonumber \\
E_{z,m,n}=&
\begin{cases}
E_{0;m,n} \cos(k_{x,m} x) \cosh(k_{x,m} y) & 0 < y < a  \\
E_{0;m,n} \frac{\cosh(k_{x,m} a)}{\sin \left[k_{y,n}(b-a)\right]}
\cos(k_{x,m} x) \sin \left[ k_{y,n}(b-y) \right] & a < y < b 
\end{cases}
\nonumber \\
\nonumber \\
H_{x,m,n}=&
\begin{cases}
\frac{i E_{0;m,n} k_z \epsilon c}{k_{x,m}} \cos(k_{x,m} x) \sinh(k_{x,m} y) & 0 < y < a \\
-\frac{i E_{0;m,n} k_z \epsilon c}{k_{x,m}} \frac{\cosh(k_{x,m} a)}
{\sin \left[ k_{y,n}(b-a) \right]}
\cos(k_{x,m} x) \cos \left[ k_{y,n}(b-y) \right] & a < y < b
\end{cases}
\nonumber \\
\nonumber \\
H_{y,m,n}=& 0
\nonumber \\
\nonumber \\
H_{z,m,n}=&
\begin{cases}
E_{0;m,n} \epsilon c \sin(k_{x,m} x) \sinh(k_{x,m} y) & 0 < y < a  \\
\frac{E_{0;m,n} k_{x,m} \epsilon c}{k_{y,n}} \frac{\cosh(k_{x,m} a)}{\sin \left[k_{y,n}(b-a)\right]}
\sin(k_{x,m} x) \cos \left[ k_{y,n}(b-y) \right] & a < y < b
\end{cases}
\end{align}
\end{widetext}

\subsection{Longitudinal Section Electric (LSE) modes}

The case of the LSE modes ($E_y =0$) can be treated in the same way
as the LSM modes. The hertzian vector electric potential 
is replaced by a vector magnetic potential $\boldsymbol{\Pi}_h$ which is
related to the fields:

\begin{eqnarray}
\label{vhp_lse1}
\mathbf{E} &=& -i \omega \mu \boldsymbol{\nabla} \times
\mathbf{\Pi_h} \\
\label{vhp_lse2}
\mathbf{H} &=& k^2 \mathbf{\Pi_h} + \boldsymbol{\nabla}
(\boldsymbol{\nabla} \cdot \mathbf{\Pi_h})
\end{eqnarray}

As in the previous case, it is convenient to factor out the
$t$ and $z$-dependencies of the $\mathbf{\Pi_h}$
and to define a scalar
function $\psi_h (x,y)$: $\mathbf{\Pi_h} =\psi_h(x,y) e^{i(\omega t-k_z z)}  \hat{\mathbf{y}}$. It is straightforward
to derive the dispersion equation for the LSE modes

\begin{equation}
\label{disp_lse}
\coth(k_x a) \cot \left[ k_y(b-a) \right] = -\frac{k_x}{k_y}, 
\end{equation}

\noindent and the expressions for the electromagnetic-field components: 

\begin{widetext}
\begin{align}
\label{Ez_LSE}
E_{x,m,n}=&
\begin{cases}
-\frac{i E_{0;m,n} k_z}{k_{x,m}} \sin(k_{x,m} x) \cosh(k_{x,m} y) & 0 < y < a \\
-\frac{i E_{0;m,n} k_z}{k_{x,m}} \frac{\cosh(k_{x,m} a)}{\sin \left[ k_{y,n}(b-a) \right]}
\sin(k_{x,m} x) \sin \left[ k_{y,n}(b-y) \right] & a < y < b
\end{cases}
\nonumber \\
\nonumber \\
E_{y,m,n}=& 0
\nonumber \\
\nonumber \\
E_{z,m,n}=&
\begin{cases}
E_{0;m,n} \cos(k_{x,m} x) \cosh(k_{x,m} y) & 0 < y < a  \\
E_{0;m,n} \frac{\cosh(k_{x,m} a)}{\sin \left[k_{y,n}b-a)\right]}
\cos(k_{x,m} x) \sin \left[ k_{y,n}(b-y) \right] & a < y < b
\end{cases}
\nonumber \\
\nonumber \\
H_{x,m,n}=&
\begin{cases}
\frac{i E_{0;m,n} k_{x,m}}{k_z \mu c} \cos(k_{x,m} x) \sinh(k_{x,m} y) & 0 < y < a \\
-\frac{i E_{0;m,n} k_{y,n}}{k_z \mu c} \frac{\cosh(k_{x,m} a)}
{\sin \left[ k_{y,n}(b-a) \right]}
\cos(k_{x,m} x) \cos \left[ k_{y,n}(b-y) \right] & a < y < b
\end{cases}
\nonumber \\
\nonumber \\
H_{y,m,n}=& 
\begin{cases}
\frac{i E_{0;m,n} (k_{x,m}^2+k_z^2)}{k_{x,m} k_z \mu c} \sin(k_{x,m} x) \cosh(k_{x,m} y)
& 0 < y < a \\
\frac{i E_{0;m,n} (k_{x,m}^2+k_z^2)}{k_{x,m} k_z \mu c} \frac{\cosh(k_{x,m} a)}
{\sin \left[ k_{y,n}(b-a) \right]}
\sin(k_{x,m} x) \cos \left[ k_{y,n}(b-y) \right] & a < y < b
\end{cases}
\nonumber \\
\nonumber \\
H_{z,m,n}=&
\begin{cases}
\frac{E_{0;m,n}}{\mu c} \sin(k_{x,m} x) \sinh(k_{x,m} y) & 0 < y < a  \\
-\frac{E_{0;m,n} k_y}{k_{x,m} \mu c} \frac{\cosh(k_{x,m} a)}{\sin \left[k_{y,n}(b-a)\right]}
\sin(k_{x,m} x) \cos \left[ k_{y,n}(b-y) \right] & a < y < b
\end{cases}
\end{align}
\end{widetext}

%\subsection{Discussion}

%For higher-order modes the solutions of both LSM and LSE dispersion
%equations tend to be equally spaced. For large values of
%$x \equiv k_y (b-a)$, the solutions of the transcendental
%equations~\ref{disp_lsm} and~\ref{disp_lse}
%approach $p \pi$ and
%$\left(p+\frac{1}{2} \right) \pi$ respectively
%where the $p \gg 1$ is an integer. 

%The dispersion equations for the LSM and LSE dipole modes is obtained by substituting $\tanh (k_{x,m} a)$ for $\coth (k_{x,m} a)$ in Equations.~\ref{disp_lsm} and~\ref{disp_lse} respectively.

%An important property of the wakefields described by Eqns.~\ref{Ez}
%and~\ref{Ez_LSE} is that the phase difference between the 
%longitudinal and transverse components is always $\pm \frac{\pi}{2}$ 
%no matter what the mode type is. This means that whenever the
%longitudinal force, $F_z = q E_z$, reaches a minimum (drive bunch)
%or a maximum (test bunch), the vertical force, $F_y = q \left(
%E_y + v B_x \right)$ is zero. The important consequence is that
%the beam breakup cannot happen for these kind of structures
%provided the electron bunch length is much shorter than the
%wavelength of the fundamental mode. 

\section{WAKEFIELD AMPLITUDES}

Although we have obtained the general expression for the electromagnetic field, the constants $E_{0;m,n}$ still remained to be determined. An often used method to find the constants $E_{0;m,n}$ in Eqns.~\ref{Ez} and~\ref{Ez_LSE}, is to evaluate the Green function of the wave equation with sources included. The latter is usually a cumbersome procedure and we instead chose to determine $E_{0;m,n}$'s  based on energy balance considerations. The electromagnetic energy stored in the DLW equals the mechanical work performed on the drive bunch. For a linear wakefield, the decelerating
field $E_d$ acting on a drive point-charge is half of the wakefield amplitude, i.e. $E_d = \frac{E_{0;m,n}}{2}$as a consequence of the fundamental wakefield theorem~\cite{Ruth}. . Suppose the drive charge 
moves over an infinitely short distance $\delta z$. The work performed by the
wakefield on the drive charge should equal the energy stored
in the field
\begin{align}
\label{en_bal}
& \sum_{m,n} \int \delta (y-y_0) \frac{E_{z;m,n} (z=vt)}{2} dxdy\delta 
z  \nonumber \\
& = \frac{1}{2} \int \left( \epsilon E^2 + \mu_0 H^2 \right) dx dy \delta z, 
\end{align}
where the integration extends over the transverse plane. It is straightforward to evaluate the full expressions
of the wakefield amplitudes in terms of the wave numbers $k_{x,m}$ and $k_{y,n}$
from Eqns.~\ref{en_bal},~\ref{Ez} and~\ref{Ez_LSE}:

\begin{widetext}
\label{E0}
\begin{equation}
E_{0; m,n}^{LSM} = \frac{1}{2 \epsilon_0} \frac{\lambda_{m}
\cosh (k_{x,m} y_0)}{\frac{\sinh(2 k_{x,m} a)}{2 k_{x,m}} + \frac{\epsilon_r 
\cosh^2(k_{x,m} a)}{\sin^2 \left[ k_{y,n}(b-a) \right]} \left\lbrace \frac{b-a}{2}
\left( 1 + \frac{\epsilon_r k_{x,m}^2}{k_{y,n}^2} \right) - 
\frac{\sin \left[2k_{y,n}(b-a) \right]}{4k_{y,n}} \left( 1-
\frac{\epsilon_r k_{x,m}^2}{k_{y,n}^2} \right) \right\rbrace} \nonumber
\end{equation}

\begin{equation}
E_{0; m,n}^{LSE} = \frac{1}{2 \epsilon_0}  \frac{\lambda_{m}
\cosh (k_{x,m} y_0)}{\frac{\sinh(2 k_{x,m} a)}{2 k_{x,m}} + \frac{
\cosh^2(k_{x,m} a)}{\sin^2 \left[ k_{y,n}(b-a) \right]} \left\lbrace \frac{b-a}{2}
\left( \epsilon_r + \frac{k_{y,n}^2}{k_{x,m}^2} \right) -
\frac{\sin \left[2k_{y,n}(b-a) \right]}{4k_{y,n}} \left( \epsilon_r -
\frac{k_{y,n}^2}{k_{x,m}^2} \right) \right\rbrace} 
\end{equation}

\end{widetext}

The field amplitudes for the dipole modes can be obtained from the
previous equations by substituting $\sinh(k_{x,m} u)$ for $\cosh(k_{x,m} u)$
where $u$ takes on the values of $a$ and $y_0$.
As expected, the wakefield amplitude scales linearly with the drive
bunch charge and inverse proportionally with the transverse size
of the structure. To this point this model is fully three-dimensional
and the only limitation stems from the assumed symmetry of the transverse
drive charge distribution with respect to the $y$ axis.

\begin{figure}[htb]
\centering
\includegraphics*[width=0.51\textwidth]{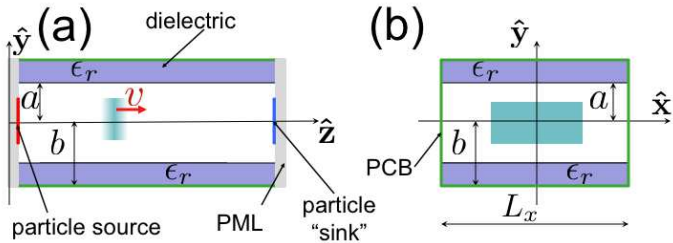}
\caption{(color online) Longitudinal (a) and transverse sections (b) of the DLW model implemented in {\sc vorpal}. The rectangular box delimited by green blocks represents the 3D computational domain used in the simulations. Grey, green and purple blocks respectively stand for perfectly-matched layer (PML), perfectly-conducting boundary (PCB), and dielectric material (with associated relative dielectric permittivity $\epsilon_r$). The cyan rectangles represent the electron bunch distribution. }
\label{vorpalmodel}
\end{figure}

\section{COMPARISON WITH three-dimensional FDTD SIMULATIONS}

To evaluate the wakefield associated to a drive bunch, an integration over the full three-dimensional continuous charge distribution must be performed. 
In practice the integration is replaced by numerical summations of discrete charge distributions similar to those described by Eq.~\ref{charge}. This process is similar to the charge discretization procedure used in standard PIC algorithms~\cite{picbook}. 
An important feature of this model is that the integration over $x$-direction is already performed through the Fourier
expansion of the charge distribution. For the charge distributions considered in this paper, only a few Fourier terms ($<10$) are 
sufficient to obtain an accurate representation of the distribution along the $x$ direction. This is significantly less than the number
of grid points in $x$-direction needed by most PIC codes to evaluate the 
3D-collective effects and external fields.

\begin{figure}[htb]
\centering
\includegraphics[width=0.41\textwidth,angle=-90]{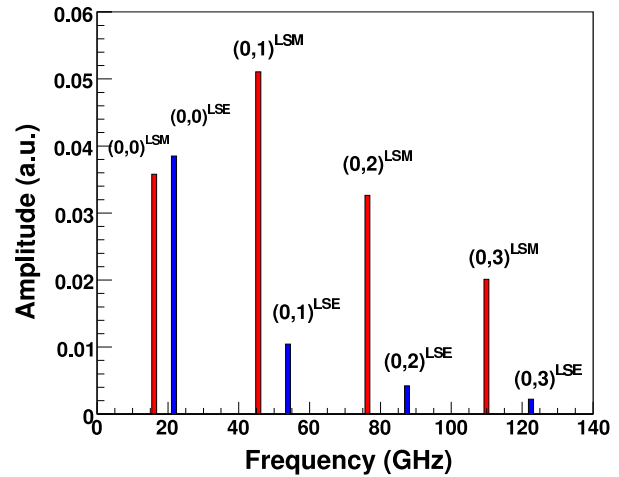}
\caption{Eigenfrequencies [$f_{m,n}\equiv\omega_{m,n}/(2\pi)$] and associated amplitudes for the mode 
induced in the DLW with parameter listed in Tab.~\ref{tab1}.  The $(m,n)$ LSM and LSE modes are respectively shown as red and blue bars .}
\label{graph0}
\end{figure}

The integration over the vertical direction is 
straightforward and in most cases, when the
drive charge distribution is also symmetric with 
respect to the horizontal axis, the
contribution of the dipole modes cancels out. 

Since the phase velocity of the wakefield is the same as the 
velocity of the drive beam, causality principle requires that
the wakefield vanishes ahead of the drive charge.
Therefore, the integration over the longitudinal direction 
extends only from the observation 
point to the actual drive charge position. The wakefield assumes the form 

\begin{equation}
\label{conv}
W(z) = \sum_{m=0,1,\cdots}\sum_{n=0,1,\cdots} \int_{z}^{\infty} \rho (z') W_{m,n} (z-z') dz', 
\end{equation}
where $W$ stands for any of the electromagnetic field components and $W_{m,n} $ are the corresponding field component associated to the LSM$_{m,n}$ and LSE$_{m,n}$ modes given respectively by  Eq.~\ref{Ez} and Eq.~\ref{Ez_LSE}.  The  summation is performed over all excited modes. \\

The results obtained from Eq.~\ref{conv}  are benchmarked against simulations performed with {\sc vorpal} a conformal FDTD (CFDTD) PIC electromagnetic solver~\cite{VORPAL}. {\sc vorpal} is a parallel, object-oriented framework for three dimensional relativistic electrostatic and electromagnetic plasma simulation. The DLW model implemented in {\sc vorpal} is fully three dimensional; see Fig.~\ref{vorpalmodel}. The model consists of  the rectangular DLW surrounded by perfectly-conducting boundaries (PCBs), The lower and upper $z$ planes are terminated by perfectly matched layers (PMLs) that significantly suppresses artificial reflections of incident radiation~\cite{Berenger}. A particle source located on the surface of the lower $z$ plane produced macroparticles  uniformly distributed in the transverse plane $(x,y)$ and following a Gaussian longitudinal distribution with total charge $Q$ described by 
\begin{eqnarray}
\label{chargedist}
\Xi(x,y,\zeta) &=& \frac{Q}{\sqrt{2\pi}\sigma_z w_x w_y} e^{-\frac{\zeta^2}{2\sigma_z^2}} H\left(\frac{w_x}{2}-|x|\right)  \nonumber \\
& & \times  H\left(\frac{w_y}{2}-|y|\right)  H\left(3\sigma_z -|\zeta|\right), 
\end{eqnarray}
where $\zeta\equiv z-vt$, $w_x$ ($w_y$) is the full width transverse horizontal (vertical) beam size, and $\sigma_z$ the longitudinal root-mean-square (rms)  length. The function $H(...)$ is the Heaviside function. The longitudinal Gaussian distribution is truncated at $\pm 3 \sigma_z$.
 Finally, a  ``particle sink" at the upper $z$ plane allows macroparticles to exit the computational domain without being scattered or creating other source of radiation.   \\

\begin{figure}[htb]
\centering
\includegraphics*[width=0.5\textwidth]{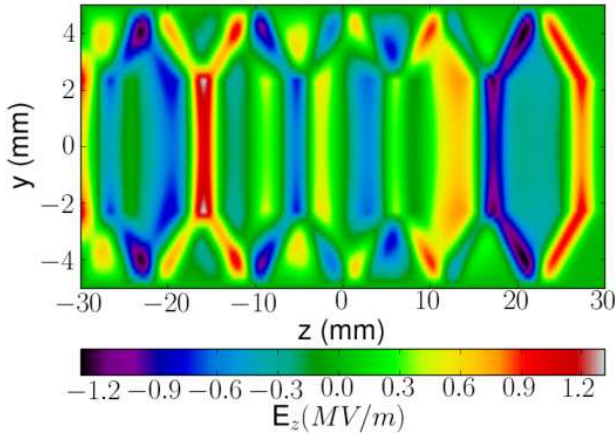}
\caption{Snapshot of the axial electric field $E_z(x=0,y,z)$ in the mid-plane of a slab DLW for the bunch and structure parameters shown in Tab.~\ref{tab1}. The center of the drive bunch is
located at $z=27$~mm and it is moving in the positive $z$-direction. The field was obtained from {\sc vorpal} simulations.}
\label{yzview}
\end{figure}

In order to precisely benchmark our model, a DLW which supports both LSM and LSE modes is chosen. The parameters of the structure and driving bunch are gathered in Tab.~\ref{tab1} and the frequency and the amplitude associated to the first few modes appear in Fig.~\ref{graph0}.   The drive-bunch energy is set to ${\cal E}=1.0$~GeV consistent with the ultra-relativistic approximation used in the analytical model.

\begin{table}[hbt]
\caption{Parameters of the DLW structure and drive bunch used for benchmarking our theoretical model with {\sc vorpal} simulations. \label{tab1} }

\begin{center}
\begin{tabular}{l c c c}\hline\hline\
parameter & symbol & value & unit \\
\hline
vacuum gap  & $a$  & 2.5 & mm  \\
height  &  $b$    & 5.0 &mm  \\
width & $L_x$  & 10.0  & mm  \\
relative permittivity & $\epsilon_r$  & 4.0  & $--$ \\
\hline
bunch energy & ${\cal E}$   & 1  & GeV  \\
bunch charge  &  $Q$    & 1.0 & nC  \\
rms bunch length  & $\sigma_z$  & 1.0 & mm  \\
bunch full width & $w_x$     & 6.0  & mm  \\
bunch full height & $w_y$   & 4.0  & mm  \\
\hline \hline
\end{tabular}
\end{center}
\end{table}

\begin{figure}[htb]
\centering
\includegraphics*[width=0.5\textwidth]{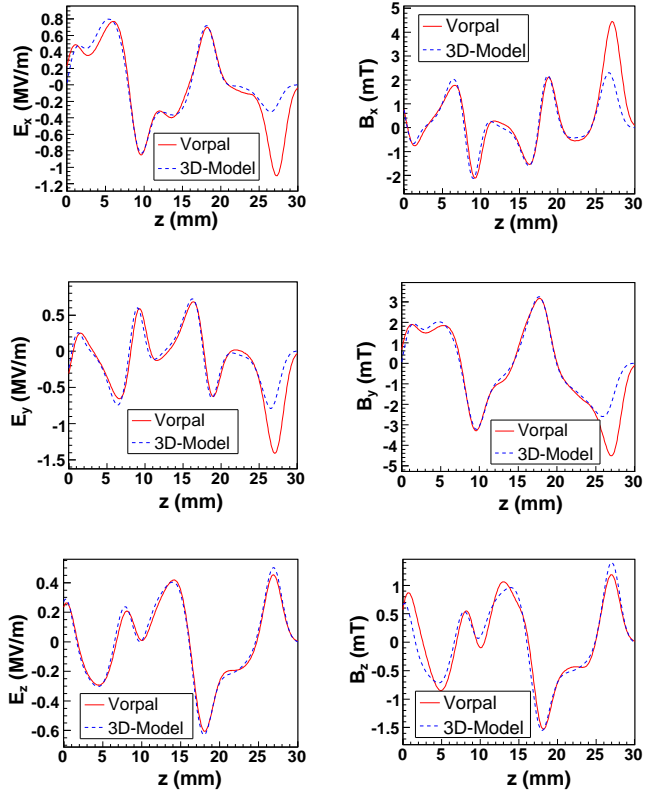}
\caption{Comparison of the electromagnetic field components calculated with our model (blue dashed line) and simulated with {\sc vorpal} (red solid lines). The fields are computed on a line parallel to the $z$ axis with transverse offset $x=3.0$~mm and $y=2.0$~mm. The center of the drive beam is at $z=27.5$~mm. The DLW and bunch parameters are the one displayed in Tab.~\ref{tab1}.}
\label{graph1}
\end{figure}

\begin{figure}[htb]
\centering
\includegraphics*[width=0.5\textwidth]{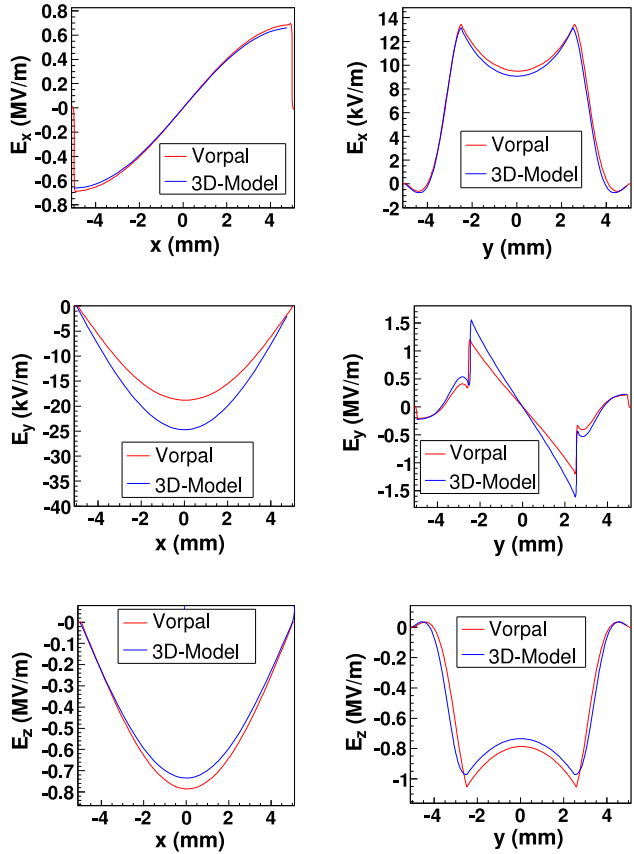}
\caption{Comparison of the electromagnetic field components calculated with our model (blue dashed line) and simulated with {\sc vorpal} (red solid lines). The fields are computed a a given longitudinal position $z=18.6$~mm. The left-column plots display the dependence on the horizontal coordinate $x$ at $y=0.043$~mm while the right-column plots show the dependence on the vertical coordinate $y$ at $x=0.043$~mm. The DLW and bunch parameters are the one displayed in Tab.~\ref{tab1}.}
\label{graph2}
\end{figure}

The longitudinal component of the electric field simulated with {\sc vorpal}  is shown in 
Fig.~\ref{yzview} as a two-dimensional projection in the $y-z$ plane.

The comparison between theoretical calculation and {\sc vorpal} simulations
are shown in Figs.~\ref{graph1} and~\ref{graph2}. The fields in 
Fig.~\ref{graph1} are evaluated as a function of the axial coordinate $z$ at a given transverse location 
($x = 3.0$~mm, $y=2.0$~mm) which corresponds to the upper-left
corner of the charge distribution. Figure~\ref{graph2} displays the electromagnetic field evaluated as a function of the horizontal [Fig.~\ref{graph2} (left)] and vertical [Fig.~\ref{graph2} (right)] transverse coordinate at a given axial location $z=18.6$~mm corresponding to the minimum $E_z$ shown in Fig.~\ref{graph1} (i.e. maximum accelerating field). All plots show a decent agreement (relative discrepancy $< 25$~\%) between our  model and {\sc vorpal} simulations. The notable disagreement observed in Fig.~\ref{graph1} for the transverse fields in the vicinity of  the driving charge (i.e. $z\simeq 27.5$~mm) is rooted in the absence of velocity fields in our  model (only radiation field contributes to the wakefield) while the {\sc vorpal} simulations include both velocity and radiation fields. In fact given the bunch charge and duration, the amplitude of the velocity field can be evaluated by convolving the charge distribution with the electric field generated by an ultra-relativistic particle $\mathbf{E}(r,\zeta)=q/(2\pi\epsilon_0) (\mathbf{r}_{\perp}/r_{\perp}^2) \delta(\zeta)$ where $e$ and $\epsilon_0$ are respectively the electronic charge and vacuum permittivity, and $\mathbf{r}_{\perp}\equiv (x,y)$ and $\zeta$ are respectively the transverse and longitudinal coordinates of the observation point referenced to the particle's location.  Such a convolution with Eq.~\ref{chargedist} yields the transverse electric field components at $(x,y,\zeta)=(w_x,w_y,0)$ to be  $\sim 1$~MV/m for the bunch parameter listed in Tab.~\ref{tab1}. The latter value is in agreement with the observed difference between the {\sc vorpal} and theoretical models; see $E_x$ and $E_y$ components in Fig.~\ref{graph2}. 

\section{THE TWO-DIMENSIONAL limit}

An interesting limiting case occurs when $L_x \gg L_y$. In this ``two-dimensional limit",  the dependence of the fields on the horizontal coordinate $x$ is weak and completely vanishes when $L_x \longrightarrow \infty$ (so that $k_x\simeq 0$ and $m=0$). In such a case the field amplitudes associated to the LSM and LSE modes are respectively 
\begin{align}
\label{2dcase1}
E_{0;0,n}^{LSM} \simeq  \frac{4 \Lambda}{a + \frac{\epsilon_r
(b-a)}{\sin^2 \left[ k_{y,n}(b-a) \right]}}
\mbox{,~and~} 
E_{0;0,n}^{LSE}  \simeq 0, 
\end{align}

\noindent where $\Lambda$ is charge per unit length in the horizontal direction. The LSE modes are suppressed and 
the latter equation is in agreement with the results of Ref.~\cite{Tremaine}. 

This limit case can be practically reached by using structures with large aspect ratios ($L_x \gg L_y$) driven 
by flat electron beams (tailored  such that $\sigma_x \gg \sigma_y$). Flat beams can be produced in photoinjectors 
by using a round-to-flat beam transformation~\cite{derbenev,brinkmann}. In such a scheme, a beam with large angular-momentum 
is produced in a photoinjector~\cite{yine}. Upon removal of the angular momentum by applying a torque with a set of skew quadrupole, 
the beam has its transverse emittance repartitioned with a tunable transverse emittances ratio~\cite{kjk}. Flat beams with transverse sizes of 
$\approx 100$~$\mu m$ and aspect ratio of $\sim 20$ have been produced, even at a relatively-low energy of $15$~MeV~\cite{Piot}.

\begin{figure}[htb]
\centering
\includegraphics*[width=0.48\textwidth]
%{Ez_a0.1_b0.3_eps4.0_sz0.05_Lx10_NML.eps}
{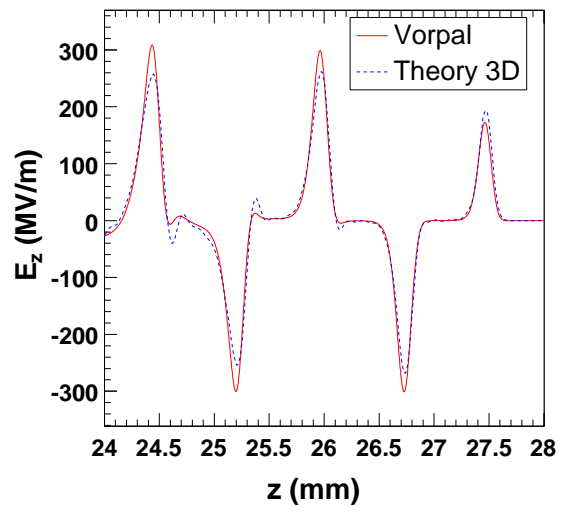}
\caption{Longitudinal electric field for a DLW with $L_x \gg L_y$ driven by a 3-nC bunch. 
The bunch has a rectangular transverse shape with full width $w_x =
3.0$~mm and $w_y = 150$~$\mu$m, and a Gaussian longitudinal distribution
with rms length $\sigma_z = 50$~$\mu$m. The theoretical model based on Eq.~\ref{2dcase1} (dashed blue line) 
is compared with {\sc vorpal} simulations (red line).}
\label{case2d}
\end{figure}

Based on the previous experience in producing flat beams and preliminary simulation 
of the Advanced Superconducting Test Accelerator (ASTA) currently in construction at
Fermilab~\cite{church}. It is reasonable to consider a  3-nC  flat beam generated from the ASTA photoinjector 
to have parameters tabulated in Tab.~\ref{tab2} when accelerated to 1~GeV. 
Considering a structure  with $a = 100$~$\mu$m, $b = 300$~$\mu$m and $\epsilon = 4.0$ would yield a 
maximum axial wakefield amplitude of $\sim 300$~MV/m; see Fig.~\ref{case2d}. The fundamental LSM 
frequency is $f_{0,0}=193$~GHz and the nearest LSE mode amplitude is $\sim 136$ times lower. 
In Fig.~\ref{case2d} the wakefield is computed with the asymptotic limit provided in Eq.~\ref{2dcase1} and  
is in excellent agreement with the FDTD simulations.   

\begin{table}[hbt]
\caption{Parameter of the DLW structure and drive bunch used for benchmarking our theoretical model with {\sc vorpal} simulations. \label{tab2} }

\begin{center}
\begin{tabular}{l c c c}\hline\hline\
parameter & symbol & value & unit \\
\hline
vacuum gap  & $a$  & 100 & $\mu$m  \\
height  &  $b$    & 300 &$\mu$m  \\
width & $L_x$  & 10.0  & mm  \\
relative permittivity & $\epsilon_r$  & 4.0  & $--$ \\
\hline
bunch energy & ${\cal E}$   & 1  & GeV  \\
bunch charge  &  $Q$    & 3.0 & nC  \\
rms bunch length  & $\sigma_z$  & 50 & $\mu$m  \\
 full (rms) bunch width &  $w_x$ ($\sigma_x$)     & 3 (0.870)  & mm  \\
 full (rms) bunch height & $w_y$ ($\sigma_y$)   & 150  (43.3) & $\mu$m  \\
\hline \hline
\end{tabular}
\end{center}
\end{table}

\section{implementation in a particle-in-cell beam dynamics program}
Although FDTD simulations provide important insight that can aid the design 
and optimization of the DLW geometry, their use to optimize a whole  linear accelerator 
would be time and CPU prohibitive. Therefore it is worthwhile to include a semi-analytical 
version of  the model developed in this paper in a well-established beam 
dynamics program {\sc impact-t}~\cite{ji}.  Our main motivation toward this choice stems from  
the availability of a wide range of beamline elements models. In addition,  {\sc impact-t} takes into 
account space-charge forces using a three-dimensional electrostatic solver. The algorithm consists in  
solving Poisson's equation in the bunch's rest frame and Lorentz-boosting the computed electrostatic 
fields in the laboratory frame. 

In a typical particle-tracking PIC program, like {\sc impact-t}, an electron bunch
is described by a set of ``macroparticles'' arranged to mimic  the bunch phase space distribution. Each macroparticle represents a large number  of electrons
(typically $10^3$ in our simulations). To evaluate the electrostatic fields 
in the bunch's rest frame the macroparticles are deposited on the cells of a 
three-dimensional grid. 
As a result of this charge deposition
algorithm, the initial charge distribution is approximated by a
set of point charges, located at the nodes of the three-dimensional
grid.
%The fields are evaluated by summing the 
%contributions of $N_{x} \times N_{y} \times N_{z}$ discrete charges
%located at the nodes of the computational grid. 
%Here, $N_{x}$, $N_{y}$
%and $N_{z}$ are the numbers of nodes in each direction of the 
%three-dimensional grid. 

The line charge density corresponding to certain $y$ and $z$-coordinates is
$\rho(x) = \sum_{x_0} q(x_0) \delta (x-x_0)$ and provided this distribution
is symmetric the Fourier coefficients $\lambda_m$ from Eq.~1 are given
by:

%In order to implement the model described in this paper it is
%convenient to replace the Fourier expansion coefficients 
%$\lambda_{m}$ from Eq.~\ref{charge} with discrete charges at 
%the grid nodes.
%Considering a charge distribution composed  
%two symmetrically located point-charges, $\rho(x) = 
%q \left[ \delta(x - x_0) + 
%\delta(x + x_0) \right]$ then 

\begin{eqnarray}
\lambda_{m} = \sum_{x_0} \frac{2q(x_{0})
\cos(k_{x,m} x_{0})}{L_{x}}.
\end{eqnarray}

In practice the linear charge distribution may not be "exactly"
symmetric. In this case the charge at an arbitrary grid point
$x_0$ is set at the average of the values given by the charge
deposition algorithm at positions $x_0$ and $-x_0$.

% In general, the $\lambda_{{m}}$'s are sums
%over the nodes with $x>0$. In practice the sums extend over all
%nodes provided that the charges located at symmetric nodes are about
%the same $q_{1} (x_0) \approx q_{2} (-x_0)$.

\begin{figure}[htb]
\centering
\includegraphics*[width=0.45\textwidth]
%{comp_VORPAL_Imp_slab_042512.eps}
{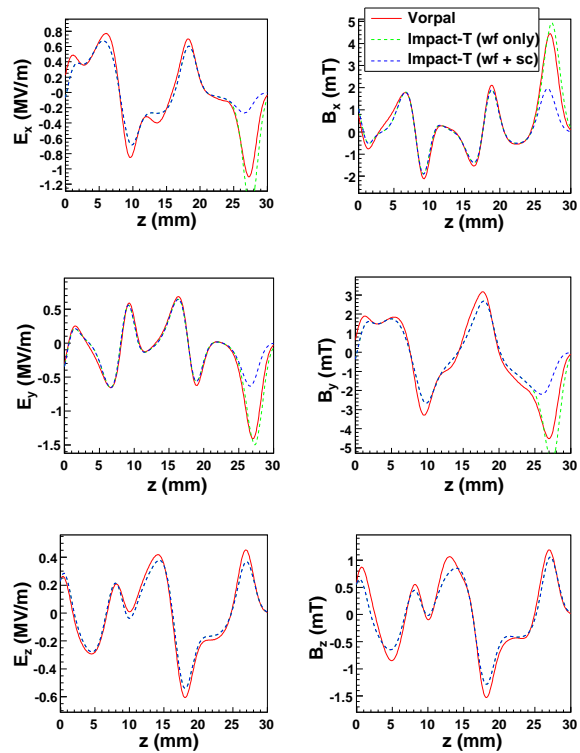}
\caption{Comparison of the electromagnetic field components obtained from {\sc impact-t} [with (green dashed line) and without (blue dashed line) accounting for space charge effects] with the fields simulated with {\sc vorpal} (red solid lines). The fields are computed on a line parallel to the $z$ axis with transverse offset $x=3.0$~mm and $y=2.0$~mm. The center of the drive beam is at $z=27.5$~mm. The DLW and bunch parameters are the one displayed in Tab.~\ref{tab1}.}
\label{casePIC}
\end{figure}

\begin{figure*}[htb]
\centering
\includegraphics*[width=0.90\textwidth]
%{daniel_particle.eps}
%{dlw_paper_fig9.eps}
{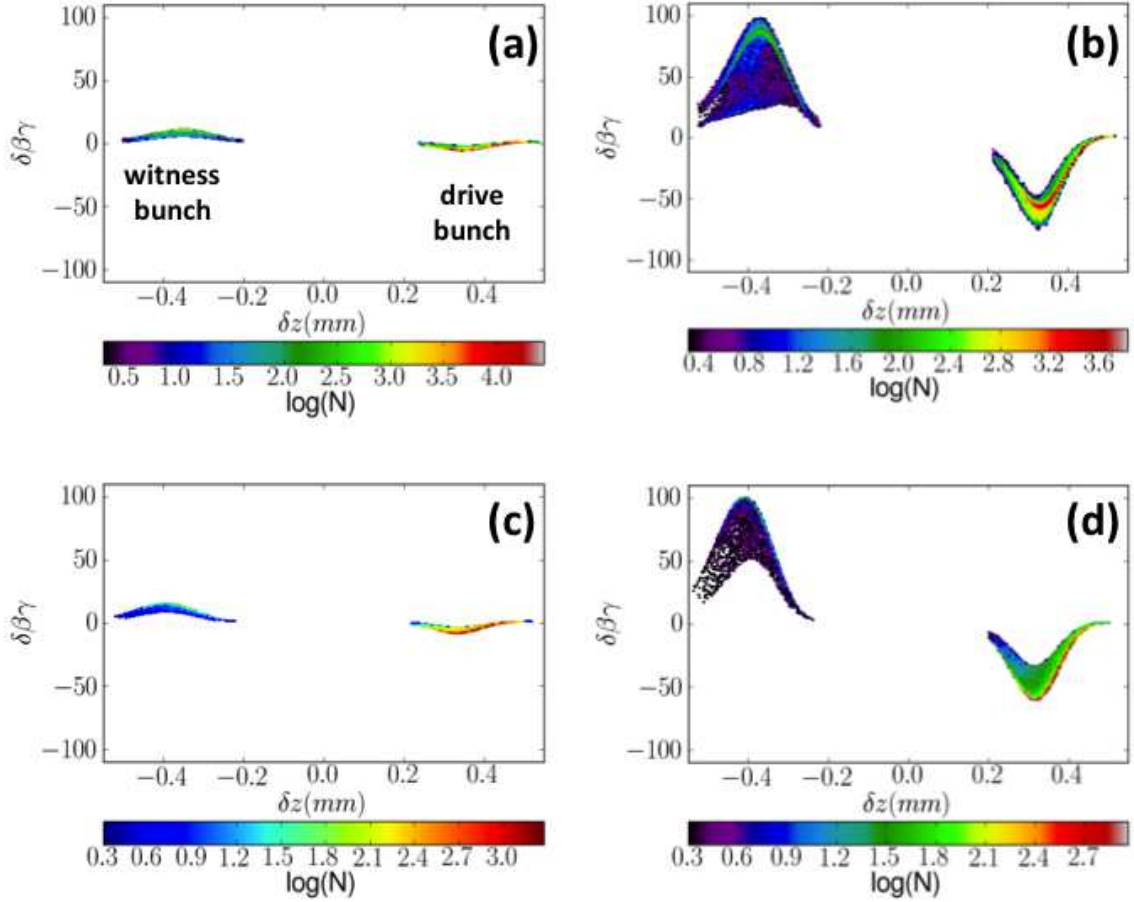}
\caption{Longitudinal phase spaces $[\delta z\equiv z-\mean{z}, \delta \beta \gamma \equiv \beta \gamma - {\cal E}/(mc)]$ snapshots as a 3-nC electron beam with parameters displayed in~\ref{tab2} propagates in a DLW structure  at $\mean{z}\simeq 2.3$~cm (a, c) and $\mean{z}\simeq 22$~cm (b, d) from the structure's entrance. Density plots (a, b) and (c, d) respectively correspond to simulations carried with {\sc vorpal} and {\sc impact-t}. The beam energy is ${\cal E}=1$~GeV. The DLW and bunch parameters are the ones appearing in Tab.~\ref{tab2}.}
\label{casePIC2}
\end{figure*}

The general expressions of the wakefields at a given position have
the following form 

\begin{equation}
\label{struct3}
W_i = \sum_{n} f_i (x,y; n) \left( \sum_{z_0 > z} \cos \left[
k_z (z_0 -z) \right] T_i (z_0; n) \right), 
\end{equation}
where $i \equiv x,y,z$, $n$ is the mode index, $f_i$'s are
geometry dependent functions of the type described in Eq.~\ref{struct1}
and

\begin{equation}
\label{Ti}
T_i (z_0;n) = \sum_{x_0, y_0} q(x_0,y_0,z_0) \cos(k_{x,m} x_0) \cosh(k_{x,m} y_0).
\end{equation}

In the latter equation the substitution $\cosh \leftrightarrow \sinh$ should be made when considering dipole modes.

The simulated electromagnetic field components are compared with {\sc vorpal} simulations in Fig.~\ref{casePIC} for 
the same case as presented in Fig.~\ref{graph1}. In the Fig.~\ref{casePIC} the {\sc impact-t} simulation are performed 
with and without activating the space-charge algorithm. When accounting for space-charge forces, {\sc impact-t} is in 
very good agreement with {\sc vorpal}. 

The convolution over the $z$-summation in Eq.~\ref{struct3} can be 
be performed with a numerical Fast Fourier Transformation (FFT).
Therefore the total number of operations needed to evaluate the wakefields
is $\propto N_{modes}
N_{x}^{2} N_{y}^{2} N_{z} \log N_{z}$. The total number of modes 
is twice the
product between the modes allowed for each of the transverse wave
numbers: $N_{modes} = 2 N_{k_{x}} N_{k_{y}}$. The factor of $2$ comes
from the inclusion of the dipole modes along with the vertically
symmetric monopole modes. For the data generated in Fig.~\ref{casePIC}, the 
{\sc impact-t} simulations are more than two orders of magnitude faster than the {\sc vorpal} ones.  \\

\begin{figure}[htb]
\centering
\includegraphics[width=0.45\textwidth]
%{energy_VORPAL_Imp.eps}
{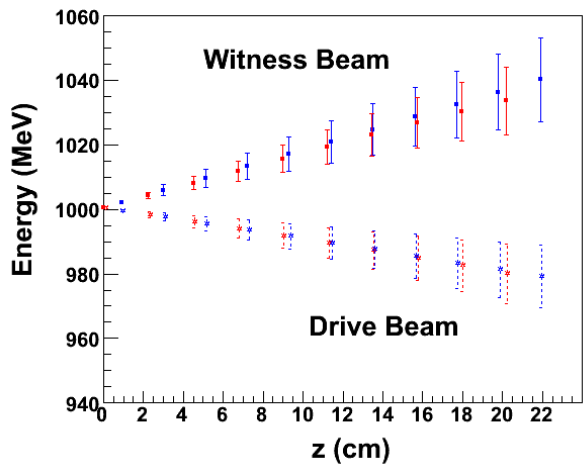}
\caption{Evolution of mean (symbols) and rms (bars) energy as the drive (top traces) and witness (bottom traces) bunches propagates along the DLW structure. The simulations are carried with {\sc vorpal} (red traces and symbols) and the modified version of {\sc impact-t} (blue traces and symbols). The beam energy is ${\cal E}=1$~GeV. The DLW and bunch parameters are the ones gathered in Tab.~\ref{tab2}. The ordinate $z=0$~cm corresponds to the DLW entrance.}
\label{EandDE}
\end{figure}

The altered version of {\sc impact-t} was used to explore a possible DLW experiment at the ASTA facility as a 1-GeV electron beam  is injected in a DLW. The beam and structure have the parameters displayed in Tab.~\ref{tab2} . For these simulations, the electron beam is an idealized cold beam without energy spread nor divergence. Figure~\ref{casePIC2} compares the longitudinal phase spaces simulated with {\sc vorpal} (top row) and {\sc impact-t} (bottom row) at two axial locations. The agreement on the phase space structure developing as this non-optimized beam propagates in the DLW is excellent. The mean and rms energies evolution for the witness and drive bunches reported in Fig.~\ref{EandDE}  confirms the quantitative agreement between the FDTD and semi-analytical models.  Both models give mean and rms energies in agreement within $\lesssim 5$~\%. 

\section{CONCLUSIONS}

In this paper we presented a three-dimensional model to evaluate wakefields
in slab-symmetric dielectric-lined waveguides. The only limitation of
this model stems from the assumed charge distribution symmetry 
with respect to the vertical axis [$\rho(x) = \rho(-x)$]. The model was 
successfully validated against three-dimensional FDTD PIC simulations performed with {\sc vorpal} 
and was implemented in the popular beam dynamics tracking program~{\sc impact-t}. 
The added capability to {\sc impact-t} enables start-to-end simulation of linear accelerators based 
on DLW accelerating structures. Furthermore, because {\sc impact-t} includes a space charge algorithm, the upgraded version 
provides a valuable tool for investigating the performances of  DLW acceleration when the dynamics of either, or 
both, of the drive and witness bunches is significantly impacted by space charge effects. The observed good agreement between 
the developed algorithm and simulations performed with the {\sc vorpal} FDTD PIC program demonstrates that our model strikes 
an appropriate balance between, efficiency, accuracy, and simplicity.  

\begin{acknowledgments}

We are thankful to J. Qiang and R. Rynes of LBNL for providing us with the sources of the {\sc impact-t} program. This work was supported by the Defense Threat Reduction Agency, Basic Research Award \# HDTRA1-10-1-0051, to Northern Illinois University. The work of D.M. is partially supported by the Fermilab Research Fellowship program under the Department of Energy contract  \# DE-AC02-07CH11359 with the Fermi Research Alliance, LLC. 

\end{acknowledgments}


\begin{thebibliography}{24}  
\bibitem{Wuensch}  
W. Wuensch, et al., Demonstration of high-gradient acceleration, Proceedings of the 2003 Particle Accelerator Conference (PAC03), Portland, Oregon, 495 (2003).

\bibitem{Aune} 
B. Aune et al.,  Phys. Rev. ST Accel. Beams {\bf 3}, 092001 (2000).

\bibitem{Leemans} 
W. Leemans et al., Nature Physics {\bf 2}, 699 (2006).

\bibitem{Melissinos} 
A. C. Melissinos, ``The energetic of particle acceleration using high intensity lasers", arXiv:physics/0410273 (2004).

\bibitem{Blumenfeld} 
I. Blumenfeld et al.., Nature  {\bf 445}, 741 (2007).

\bibitem{RosenPWFA} 
 J. B. Rosenzweig, A. M. Cook, A. Scott, M. C. Thompson, R. B. Yoder, Phys. Rev. Lett.  {\bf 95}, 195002 (2005).

\bibitem{Voss}
G. A. Voss and T. Weiland, ``Particle Acceleration by Wake Field", DESY report M-82-10, April 1982; W. Bialowons et al., ``Computer Simulations of the wakefield transformer experiment at DESY", Proceedings of the 1988 European Particle Accelerator Conference (EPAC 1988), Rome, Italy, 902 (1988).

\bibitem{Chang}
C.~T.~M.~Chang, and J.~W.~Dawson, J. Appl. Phys., {\bf {41}}, 4493 (1970).

\bibitem{Rosing}
M.~Rosing, and W.~Gai, Phys. Rev. D {\bf {42}}, 1829 (1990).

\bibitem{Ng}
K-. Y. Ng,  Phys. Rev. D  {\bf 42}, 1819 (1990).

\bibitem{Tremaine}
A.~Tremaine,  J.~Rosenzweig, and P.~Schoessow, Phys. Rev. E {\bf {56}}, No. 6, 7204 (1997).

\bibitem{Park}
S.~Y.~Park, and J.~L.~Hirshfield, Phys. Rev. E {\bf {62}}, 1266 (2000).

\bibitem{Thompson}
M.~C.~Thompson, {\it et al.}, Phys. Rev. Lett., {\bf {100}}, 214801 (2008).

\bibitem{GaiAAC10}
W. Gai, ``Consideration for a dielectric-based two-beam-accelerator linear collider", Proceedings of 2010 International Particle Accelerator Conference (IPAC10), Kyoto, Japan, 3428 (2010).

%\bibitem{ZolentsPriv}
%A. Zholents, J. Power, organizer worksgop... April 21-22 (2011). 
\bibitem{ZolentsPriv}
C. Jing, P. Schoessow, A. Kanareykin, J.G. Power, R. Lindberg and P. Piot, "A compact soft X-ray free-electron laser facility based on a dielectric wakefield accelerator". to appear in the proceedings of the ICFA future light source workshop (FLS12), March 5-9, 2012 Newport News, VA  (2012).

\bibitem{Bolotvskii}
B.~M.~Bolotvskii, Usp. Fiz. Nauk., {\bf {75}}, 295 (1961) [Sov. Phys.
Usp. {\bf {4}}, 781 (1962)].

\bibitem{Liu}
W.~Liu, and W.~Gai, Phys. Rev. ST Accel. Beams {\bf {12}}, 051301 (2009).

\bibitem{Sotnikov}
G.~V.~Sotnikov, T.~C.~Marshall, and J.~L.~Hirshfield, 
Phys. Rev. ST Accel. Beams {\bf {12}}, 061302 (2009). 

\bibitem{Power}
J.~G.~Power, W.~Gai, and P.~Schoessow, Phys. Rev. E {\bf {60}},
6061 (1999).

\bibitem{Gai}
W.~Gai, R.~Konecny, and J.~Simpson, {\it Proceedings of the 2007 Particle Accelerator Conference} (PAC07), 636 (1997).

\bibitem{Fang}
J.~M.~Fang, {\it et al.}, {\it Proceedings of the 1999 Particle Accelerator Conference} (PAC99), 3627 (1999).

\bibitem{Jing1}
C.~Jing, A.~Kanareykin, J.~G.~Power, M.~Conde, Z.~Yusof, P.~Schoessow,
and W.~Gai, Phys. Rev. Lett. {\bf {98}}, 144801 (2007).

\bibitem{Xiao}
L.~Xiao, W.~Gai, and X.~Sun, Phys. Rev. E {\bf {65}}, 016505 (2001).

\bibitem{Jing2}
C.~Jing, {\it et al.}, Phys. Rev. E
{\bf {68}}, 016502 (2003).

\bibitem{Wang}
C.~Wang, and J.~L.~Hirshfield, Phys. Rev. ST Accel. Beams {\bf {9}},
031301 (2006).

\bibitem{Brinkmann}
R.~Brinkmann, Y.~Derbenev, K.~Fl\"ottmann, Phys. Rev. ST Accel. Beams
{\bf {4}}, 053501 (2001).

\bibitem{Piot}
P.~Piot, Y.-E~Sun, and K.-J.~Kim, Phys. Rev. ST Accel. Beams
{\bf {9}}, 031001 (2006).

\bibitem{VORPAL}
C. Nieter and J. R. Cary, J. Comp. Phys., {\bf 196}, 448 (2004); see also http://www.txcorp.com/

\bibitem{Collin}
R.~E.~Collin, "Field Theory of Guided Waves", (1960).

\bibitem{Ruth}
R.~Ruth, {\it et al.}, Part. Accel. {\bf {17}}, 171 (1985);
K.~Bane, {\it et al.}, IEEE Trans. Nucl. Sci. {\bf {32}}, 3524 (1985).


\bibitem{Berenger}
J.~Berenger, Journal of Comp. Phys., {\bf {114}}, 185 (1994).

\bibitem{picbook} 
R. W. Hockney and J. W. Eastwood, {\em Computer Simulation using Particles}, Adam Hilger, Bristol and New York (1988). 

\bibitem{derbenev}
A. Burov, S. Nagaitsev, Ya. Derbenev,  {\em Phys. Rev.}  {\bf E 66}, 016503 (2002).

\bibitem{brinkmann}
R. Brinkmann, Ya. Derbenev and K. Fl\"ottmann,   {\em Phys. Rev. ST Accel. Beams}  {\bf 4}, 053501 (2001).

\bibitem{yine}
Y.-E Sun, {\em et al.},  {\em Phys. Rev. ST Accel. Beams} {\bf 7}, 123501 (2004).

\bibitem{kjk}
K.-J. Kim, {\em Phys. Rev. ST Accel. Beams}  {\bf  6}, 104002 (2003).

\bibitem{church}
M. Church, S. Nagaitsev and P. Piot, {\it Proceedings of the 2007 Particle Accelerator Conference} (PAC07), 2942 (2007).


\bibitem{ji} 
J. Qiang, S. Lidia, R. D. Ryne, and C. Limborg-Deprey, {Phys. Rev. ST Accel. Beams} {\bf 9}, 044204 (2006); J. Qiang, R. Ryne, S. Habib, and V. Decyk, {J. Comp. Phys.} {\bf 163}, 434 (2000).

\end{thebibliography}
\end{document}